\documentclass[aps,pre,
groupedaddress,superscriptaddress]{revtex4}

\usepackage{graphicx}
\usepackage{bm}
\newcommand{\gm}{\gamma_{<}}
\newcommand{\gp}{\gamma_{>}}

\begin{document}
\title{Formation of Large-Amplitude Low-Frequency Waves in\\ Capillary Turbulence on Superfluid He-II}

\author{Leonid V Abdurakhimov}
\affiliation{Institute of Solid State  Physics RAS, Chernogolovka, Moscow region, 142432, Russia}
\affiliation{Okinawa Institute of Science and Technology, Okinawa,  904-0495, Japan (present address)}

\author{German V Kolmakov}
\affiliation{New York City College of Technology, the City University of New York, Brooklyn, NY 11201,~USA} 
 
\author{Aleksander A Levchenko}
\affiliation{Institute of Solid State  Physics RAS, Chernogolovka, Moscow region, 142432, Russia} 
 
\author{Yuri V Lvov}
\affiliation{Department of Mathematical Sciences, Rensselaer Polytechnic Institute, Troy, NY 12180,~USA}  
  
\author{Igor~A~Remizov}
\affiliation{Institute of Solid State  Physics RAS, Chernogolovka, Moscow region, 142432, Russia}

\begin{abstract}
The results of experimental and theoretical studies of the parametric decay instability of capillary waves 
on the surface of superfluid helium He-II are reported. It is demonstrated that in a system of turbulent 
capillary waves low-frequency waves are generated along with the direct
Kolmogorov-Zakharov cascade of capillary turbulence. The effects of low-frequency damping and
the discreteness of the wave spectrum are discussed.
\end{abstract}

\maketitle

\section{Introduction}
 Highly developed hydrodynamic turbulence has provided a fascinating
 challenge for engineers, physicists and mathematicians for over two
 hundred years.
Turbulence appears in numerous systems ranging from planetary waves in
the earth's atmosphere to jet streams \cite{Smith:05,Landa:04}. Fully
developed hydrodynamic turbulence can be formed by interacting
vortices \cite{Frisch:95}. Turbulence may also appear in a system of
nonlinearly interacting waves \cite{Zakharov:92,Nazarenko:11}, which is referred to
as {\it wave turbulence}. 
  The concept of wave turbulence originated from Peierls'
 work \cite{Peierls:29} on anharmonic crystals.  Wave turbulence is
 manifested on planetary and interstellar scales, in the earth's
 magnetosphere and its coupling with the solar wind \cite{Southwood:78},
 in shock propagation in Saturn's bow~\cite{Scarf:81}, and in
 interstellar plasmas \cite{Bisnovatyi:95}.  Fascinating atmospheric
 phenomena, such as auroras in the high-latitude regions of the earth,
 are caused by ion wave turbulence in magnetic flux
 tubes~\cite{Southwood:78}. Wave turbulence also provides deep
 connections between the classical and quantum worlds; wave turbulence
 can result in kinetic condensation of classical waves \cite{Sun:12},
 which is similar in many ways to Bose-Einstein condensation  in
 quantum systems such as ultra-cold atoms \cite{Seman:11}.

Wave turbulence is much easier to understand than hydrodynamic
turbulence, because it is appropriate when the building blocks of a
system are linear waves that admit  analytical descriptions.  When
the nonlinear interactions between waves may be treated as weak
perturbations,  the statistics of the system become tractable
analytically. This allows one to derive the closed equation for the
spectral energy density of the waves, called the {\it kinetic
  equation}.  The kinetic equation for interacting waves has a
steady-state scale invariant (power-law) solution that describes a
constant flux of energy towards smaller scales. Such a power law
spectrum can be viewed as the wave analog of the Kolmogorov spectrum
of hydrodynamic turbulence~\cite{Frisch:95} and is referred to as the
{\it Kolmogorov-Zakharov} (KZ) spectrum of wave
turbulence \cite{Zakharov:92,Nazarenko:11}.

Surface capillary waves are the ripples that are created by a light
breeze on the surface of a pond.  They are short waves for which  surface tension is
the primary restoring force.  The dispersion relation between the wave
frequency $\omega$ and the wave number $k$ for capillary waves is
\begin{equation}
\omega(k) = \sqrt{\alpha k^3 \over \rho}, \label{eq:w}
\end{equation}
where $\alpha$ is the surface tension and $\rho$ is the fluid density.
For water, the characteristic wave length of capillary waves is less
than $\lambda = 2 \pi \sqrt{\alpha /\rho g} \approx 1.7$ cm (where $g$ is
the acceleration due to gravity).   Capillary waves  are a beautiful canonical
example of a wave-turbulent system with weak nonlinear wave
interactions.   The Kolmogorov-Zakharov spectrum
of capillary waves corresponds to the direct cascade, or flux of wave
energy, from low to high wave frequencies.  The existence and features of the KZ
spectrum for capillary waves is well established through both
experiments and
theory \cite{Zakharov:67,Pushkarev:96,Henry:00,Brazhnikov:02a,Kolmakov:04,Abdurakhimov:09}.

We study capillary waves on the surface of superfluid helium.  This
system provides an ideal testbed for studying nonlinear wave dynamics
due to its extremely low viscosity and the possibility of driving the
fluid surface directly by an oscillating electric field, virtually
excluding the excitation of bulk modes \cite{Brazhnikov:02a}. Previous
experiments with waves on quantum fluids (liquid helium and hydrogen)
allowed detailed study of the steady-state and decaying direct cascade
of capillary turbulence \cite{Kolmakov:04,Abdurakhimov:09},
modification of the turbulent spectrum by applied low-frequency
driving \cite{Brazhnikov:05} and the turbulent bottleneck phenomena in
the high-frequency spectral domain \cite{Abdurakhimov:10}.

In this paper, we show that under certain conditions low-frequency waves on the fluid surface
with frequencies lower than the driving frequency can be created in addition to the direct  
Kolmogorov-Zakharov cascade. In what follows we present our findings and discuss the mechanisms 
responsible for the low-frequency wave generation.

\begin{figure}[t] 
\begin{center}
\includegraphics[width=8cm]{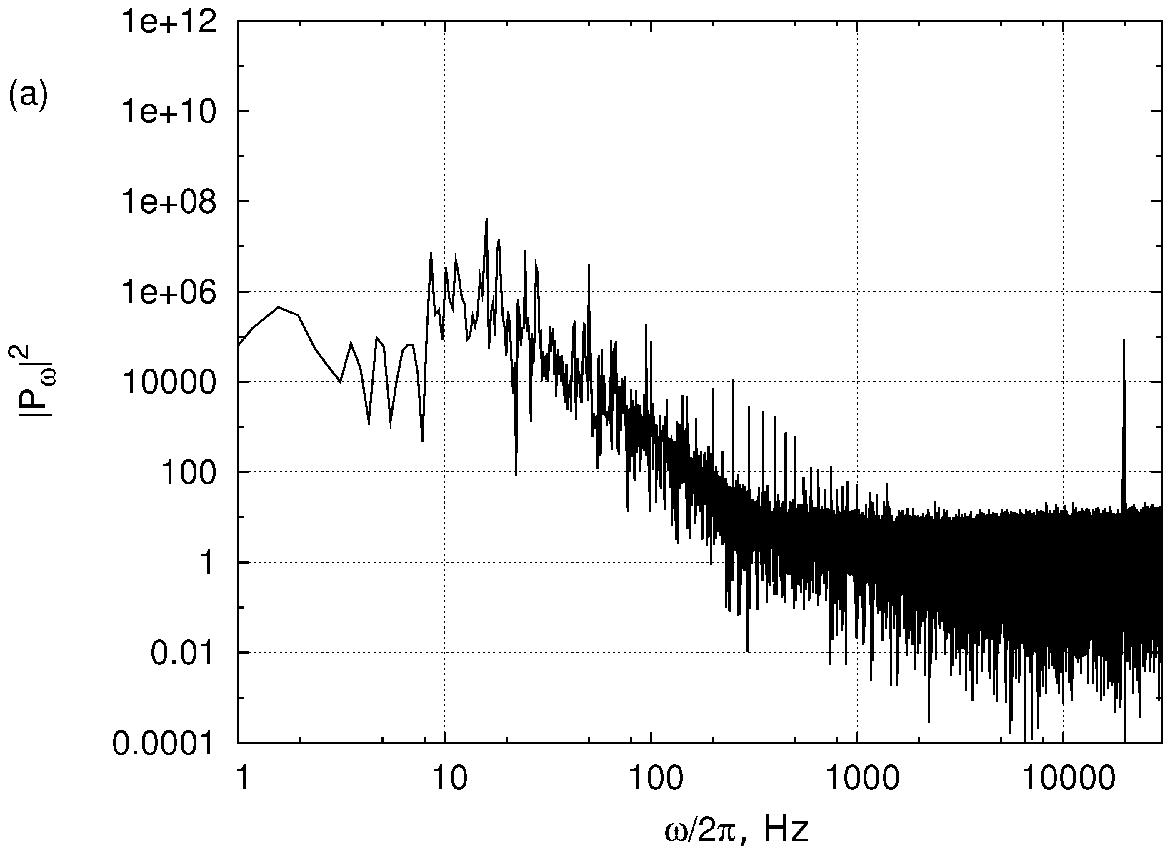}\\
\includegraphics[width=8cm]{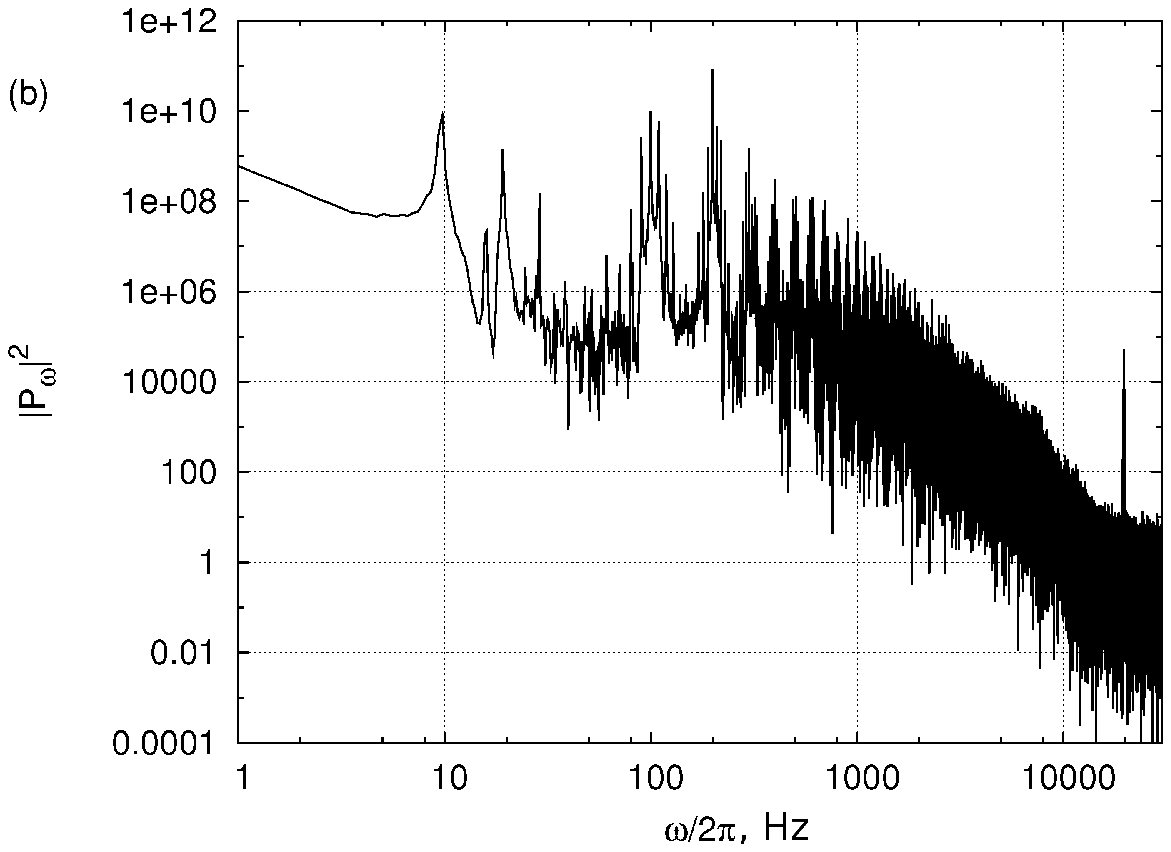}\\
\includegraphics[width=8cm]{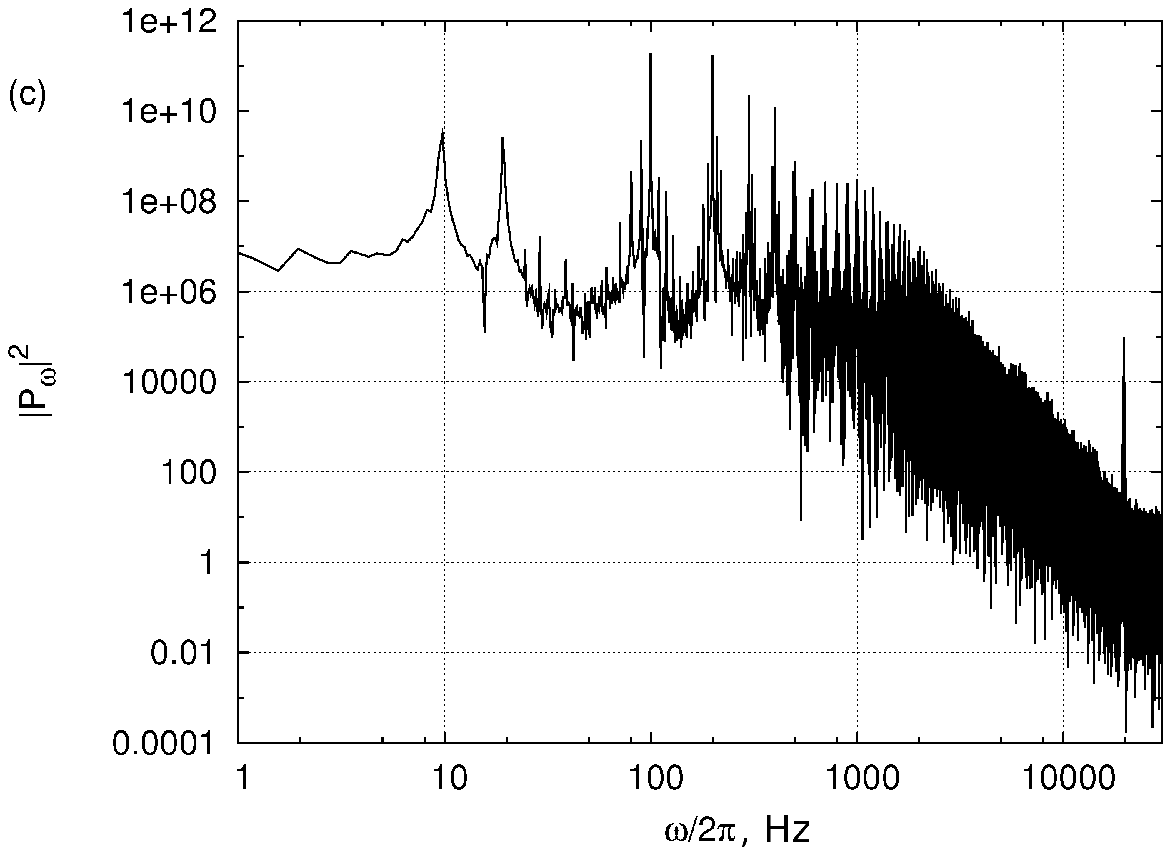}
\caption{\label{fig:1} Long-time evolution of the spectrum of capillary turbulence. 
The spectra are shown for the moments (a) $t=1.31$ s, (b) 53.74 s and (c) 163.84 s after the driving force is turned on.
Driving is applied at  a frequency $\omega_d/2\pi = 199$ Hz, the driving amplitude is $U_d=97$ V.
Formation of low-frequency harmonics with $\omega < \omega_d$ are clearly seen in figures (b) and (c).
The squared amplitudes of the harmonics $|P_{\omega}|^2$ are shown in arbitrary units.
}
\end{center}
\end{figure}

\section{Experimental observations}

 The experimental
arrangements were similar to those in our previous experiments with
superfluid helium and liquid hydrogen 
\cite{Kolmakov:04,Abdurakhimov:09}.
In our experiments, helium $^4$He was condensed into a cylindrical cup formed by the
bottom capacitor plate and a guard ring and was positioned in a helium
cryostat.  
The cup has inner radius $R=30$ mm and depth $4$ mm.  The
experiments were conducted at temperature $T=1.7$ K of the superfluid
liquid.  
The free surface of the liquid was positively charged as
the result of $\beta$--particle emission from the radioactive plate
located in the bulk liquid. Oscillations of the liquid surface were
excited by application of an AC voltage $U(t) = U_d \sin(\omega_d t)$
to the lower capacitor's plate in addition to the constant voltage.  Oscillations of the fluid surface
elevation $\zeta(\bm{r},t)$ were detected through variations of the
power $P(t)$ of a laser beam reflected from the surface.  (Here, $t$ is time and $\bm{r}$ is
two-dimensional coordinates in the surface plane). 
The power $P(t)$ was measured with a photodetector and
sampled with an analogue-to-digital converter. 
The capillary wave
power spectrum $\zeta_{\omega} \propto P_{\omega}$ was calculated via
the time Fourier transform of the signal $P(t)$ \cite{Brazhnikov:02a}.
The measurements of wave damping in the cell showed that the quality
factor at low frequencies $\omega < \omega_d$ is $Q\sim 10^3$. The
finite size of the cell results in the discrete wave number spectrum.
 The capillary-to-gravity wave transition on the surface of superfluid
helium is at frequency $\sim 30$ Hz. 
The surface oscillations with frequencies $\sim 30$ Hz are gravity-capillary waves, 
for which the restoring force is caused by both the capillary force and gravity.
However, this frequency decreases at high pulling external electric fields 
normal to the surface \cite{Brazhnikov:02a}.
 The finite depth of the waves
only influences the wave dispersion $\omega=\omega({k})$ at low
frequencies $\omega < 10$ Hz. The capillary wave length for liquid helium 
is $\lambda = 0.17$ cm at $T=4.2$ K and increases to 0.3 cm for $T=1.7$ K.

Figure \ref{fig:1} shows formation of a wave-turbulent spectrum 
after switching on the driving force at the moment of time $t=0$.
The driving frequency is $\omega_d/2\pi = 199$ Hz. 
In Fig.\ \ref{fig:1}a, waves with the frequency $\omega_d$ and its high-frequency harmonics
begin to form on the noisy background. The latter is caused by  mechanical vibrations of the installation.
In Fig.\ \ref{fig:1}b,c the Kolmogorov-Zakharov direct cascade of capillary turbulence is formed at $\omega>\omega_d$.
It corresponds to transfer of the wave energy in a cascade-like manner towards higher frequency, in full 
agreement with our previous observations \cite{Brazhnikov:02a,Kolmakov:04,Brazhnikov:05,Abdurakhimov:09,Abdurakhimov:10}.
At  frequencies $\omega/2\pi \sim 20$ kHz, the cascade is cut due to viscous damping in the fluid.
The peak at $20.5$ kHz is caused by the  noise of the He-Ne laser output power.

It is seen in Fig.\ \ref{fig:1}b,c that {\it low-frequency} waves with $\omega < \omega_d$ are formed at 
large time $t \geq 50$ s after the drive is turned on. 
Specifically, the squared amplitude of the harmonics at $\omega_d/2\pi = 99$~Hz is $|P_{\omega}|^2 \approx 8.3 \times 10^9$ a.u
at $t=53.74$ s (Fig.\ \ref{fig:1}b), and it reaches   $|P_{\omega}|^2 \approx 2.2 \times 10^{11}$ a.u.
at $t=163.84$ s (Fig.\ \ref{fig:1}c). It is remarkable that in Fig.\ \ref{fig:1}c, the amplitude of the subharmonics 
is larger than that for the wave at the drive frequency, which is $|P_{\omega}|^2 \approx 1.5 \times 10^{11}$ a.u.
The amplitudes of the low-frequency harmonics at $\omega/2\pi \approx 10$ Hz shown in Fig.\ \ref{fig:1}c
are larger than the amplitudes of the high-frequency waves
in the direct cascade with  $\omega/2\pi \geq 500$ Hz. It is worth noting that formation of the low-frequency harmonics 
on a turbulent capillary-wave background has not yet been reported in the literature.

Generation of low-frequency waves shown in Fig.\ \ref{fig:1} can be attributed to the development of the 
decay instability of capillary waves at enough large wave amplitudes.  The origin of this instability 
is in the modulation of a nearly periodic wave due to nonlinearity. This specific mechanism 
was earlier proposed to account for the creation of giant low-frequency waves on the 
water surface \cite{Dyachenko:05,Onorato:01}. It was also demonstrated that for waves on 
an ideal fluid with no damping, development of the decay instability should result in
formation of a thermodynamic-equilibrium wave distribution $|P_{\omega}|^2 \propto \Theta/\omega(k)$
where $\Theta$ is the effective temperature \cite{Balkovsky:95}. However, it is seen in Fig.\ \ref{fig:1}
that the observed spectrum at $\omega < \omega_d$ significantly differs from the proposed theoretical 
equilibrium spectrum.

\section{Numerical simulations}
 To understand the formation of large-amplitude low-frequency waves, we
performed numerical modeling of the wave dynamics in the cylindrical
cell with external driving and viscous damping.  In the simulations,
the deviation of
the surface from the equilibrium flat state is expressed by time-dependent
amplitudes $a_k(t)$ of the normal modes \cite{Zakharov:67}. We assume
angular symmetry of the surface, so its deviation for capillary waves
is
$\zeta(r,t) = \sum_k \sqrt{ k / 2 \omega_k \rho A J_0(\beta_i)^2}
(a_k(t) + a_k^*(t))J_0(kr)$, 
where $r$ is the distance from the center of the cell, $J_0(x)$ is the
Bessel function of the zero order, $A$ is the free-surface area,
$\omega(k)$ is the linear dispersion relation (\ref{eq:w}), $k \equiv
k_n = \beta_n /R$ is the radial wave number, $n >0$ is an integer
index labeling the resonant radial modes, and $\beta_n$ is the $n$th
zero of the first order Bessel function $J_1(\beta_n)=0$.  In the
simulations $r$ is measured in the capillary length scale $\lambda_c = ({\alpha /\rho g})^{1/2}$,
and time $t$ is measured in the units of $t_c = \omega_c^{-1}$, where $\omega_c=(\rho g^3/\alpha)^{1/4}$. 
The driving force is applied at a given radial mode $k_d$.  Due to angular
isotropy, we utilize the angle-averaged dynamical equation for
$a_k(t)$ \cite{Pushkarev:96},
\begin{eqnarray} % two column version
  {d a_{k}(t) \over dt}  & = &   - i \sum_{k_1,k_2} V_{k, k_1,k_2}\,  D _{k, k_1, k_2} \,
    a_{k_1}(t)  a_{k_2}(t)    e ^{i(\omega(k) - \omega({k_1})  - \omega({k_2}))t}\nonumber \\
   &  -  &   2i  \sum_{k_1,k_2} V_{k_1, k, k_2}^* \, D _{k_1, k, k_2} \, a_{k_1}(t) a_{k_2}^* (t)     e ^{i(\omega(k) + \omega({k_2}) - \omega({k_1}))t}   
     -   \gamma(\omega({k})) a_{\bm{k}}(t).
    \label{eq:3w}
\end{eqnarray}
The coupling coefficient ${V}_{k, k_1, k_2}$ characterizes the
interaction strength between waves with wave numbers $k$, $k_1$ and
$k_2$; instead of taking the exact value for capillary waves, we model
it by ${V}_{k, k_1, k_2} = \epsilon \sqrt{\omega(k) \omega({k_1})
  \omega({k_2})}$ \cite{Zakharov:92}. Star denotes complex conjugate,
$i$ stands for the imaginary unit, and
$D_{k_1, k, k_2} = 1/2 \pi \Delta (k, k_1, k_2)$,
where $ \Delta (k, k_1, k_2) $ is the area of the triangle with sides
$k$, $k_1$, and $k_2$. We consider $n_{\rm max}=100$ radial modes.
The dimensionless factor $\epsilon \ll 1$ characterizes nonlinearity
of the system and is of the order of the maximum surface slope with
respect to the horizontal \cite{Zakharov:67}. We set $\epsilon
=10^{-2}$ as a representative value \cite{Brazhnikov:02a}. Due to the
small nonlinearity, we only retain three-wave interactions in
Eq.\ (\ref{eq:3w}); the inclusion of four-wave scattering requires
special consideration \cite{During:09} and is deferred to future
studies.

We chose numerical
parameters that are representative of our experimental setup and
determined the steady-state wave spectrum from our model.
Specifically, we drive the system in the middle of our numerical
spectral range, at the 50th resonant frequency of the cell.  We also
add wave damping at both high and low frequencies, to mimic the
physical effects that remove  energy from the system. Specifically,
we  model the wave damping coefficient as
\begin{equation}
 \gamma(\omega) = \gm(\omega) + \gp(\omega), \label{eq:Damping} 
\end{equation}
which is the sum of damping at low frequencies below the 10th
resonance in the cell, with $\gm(\omega) = \gamma_{1}g_<(\omega)$, as well
as damping at high frequencies above the 80th resonance with 
$\gp(\omega) = \gamma_{0}g_>(\omega)$.
Low-frequency damping is the result of viscous drag at the cell's
bottom \cite{Christiansen:95}, and high-frequency damping models the
energy loss due to bulk viscosity in the fluid \cite{Frisch:95}.
The range of wave frequencies between the 10th and 80th resonant
frequencies can be considered as a ``numerical inertial interval'', in
which damping is absent.

\begin{figure}[t] 
\begin{center}
\includegraphics[width=8cm]{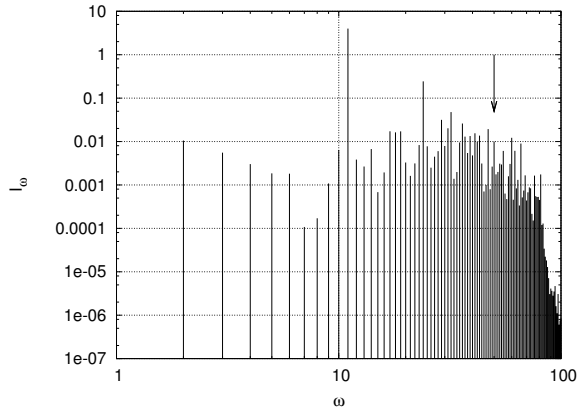}
\caption{\label{fig:2} Numerical steady-state spectrum of capillary turbulence in a cylindrical cell of
 a radius $R=30\lambda_c$. The frequency is expressed in the units of $\omega_c$. 
The waves are driven at the frequency $\omega_d$  of the 50th resonant frequency (labeled by an arrow).
The low-frequency damping coefficient is $\gamma_1 = 0.5 \gamma_0$.
It is seen that large-amplitude low-frequency waves with $\omega<\omega_d$ are formed in the system.} 
\end{center}
\end{figure}

Driving was at the 50th mode by fixing the wave amplitude $a_d \equiv
|a_{k_d}(t)|$ at a given value set; in the present simulations as
$(\lambda_c^{7/2} \omega_c \rho)^{-1/2} a_d = 0.1$.  The dimensionless
damping factor at high frequencies was set $\gamma_0 = 5\times
10^{-2}\omega_c$.  We apply damping at high resonant numbers $n>
n_{>}=80$ for which we set $ g_>(n) = (n - n_>)^2 / (n_{\rm max} -
n_>)^2$, and $ g_>(n) = 0$ for $n \leq n_>$.  To model waves on a
fluid layer of finite depth, we also apply damping at low resonant
numbers $n<n_< = 10$ as follows $g_<(n) = (n_< - n) / n_{<}$, and
$g_<(n) =0$ for $n \geq n_<$.  To calculate the dependence of $a_k(t)$
on time $t$, we integrated Eq.\ (\ref{eq:3w}) until the system reached
the steady state.  We found numerical convergence and energy
conservation with $10^{-7}$ numerical accuracy.  The normalized wave
spectrum is calculated as the time-averaged $N(k) = \langle
|a_k(t)|^2 \rangle$.  To facilitate the comparison of the simulation
results with the experimental data, the spectrum $I_{\omega} \equiv N(k)$ is expressed
as a function of the wave frequency $\omega$ via the relation
$k=k(\omega)$ from the inverse of the dispersion relation Eq. (\ref{eq:w}).

The results of the simulations for $\gamma_1 = 0.5 \gamma_0$ are shown in Fig.\ \ref{fig:2}.
It is evident that the harmonics with a frequency $\omega< \omega_d$ are formed. The amplitudes
of the low-frequency harmonics in the spectral range $0.2 \omega_d < \omega < \omega_d$
are of the  order of or larger than the amplitude at the driving frequency. This result is in qualitative agreement with 
the result of the observation shown in Fig.\ \ref{fig:1}b,c. In particular, it is
seen in Fig.\ \ref{fig:2} that the numerical spectrum $I_{\omega}$  differs from a power-like equilibrium  spectrum 
predicted for waves on an ideal fluid. The reason of this discrepancy can lie in finite damping at low frequencies
in our observations and in the numerical model. The deviation of the low-frequency spectrum can also be caused by 
the discreteness of the spectrum of resonant waves at low frequencies in a cell of finite size, 
in agreement with results of Refs.~\cite{Pushkarev:00,Lvov:10,Kartashova:10}.

\section{Conclusions}
We demonstrated that if the capillary waves on a superfluid helium surface are driven at high enough frequency, large-amplitude low-frequency waves are created in addition to  Kolmogorov-Zakharov cascade of capillary turbulence. We infer that
the reason for the low-frequency wave generation is the decay instability of capillary waves. This mechanism is previously known 
to be responsible for the inverse cascade of gravity waves on the ocean surface. The observed spectrum of low-frequency waves
differs from a pure thermal-equilibrium distribution predicted for waves on an ideal fluid. This discrepancy can be caused by the effects of 
viscous damping and by a restricted geometry of the cell. Our experimental findings are in agreement with the results of the numerical simulations
based on the wave dynamic equations.

\section*{Acknowledgments}
The authors are grateful to Prof.\  Leonid P.\ Mezhov-Deglin and
Prof.\ William L. Siegmann for valuable discussions.  L.V.A, A.A.L and I.A.R. are grateful 
to the Russian Science Foundation, grant \#14-22-00259. 
G.V.K. gratefully
acknowledges support from the Professional Staff Congress --  City University of New York award 
\# 67143-00 45.  
 Yu.V.L. is grateful for support to ONR, award
\#N000141210280. 
The authors are grateful  to the Center  for Theoretical Physics of the 
New York City College of Technology for providing computational resources.
This work is supported in part by  Army Research Office, grant \#64775-PH-REP.

%\section*{References}
%\bibliographystyle{iopart-num}
%\bibliography{../../lowtempgk}

\end{document}